\input{aipcheck}

\def\lsim{\mathrel{\rlap{\lower4pt\hbox{\hskip1pt$\sim$}}
    \raise1pt\hbox{$<$}}}         
\def\gsim{\mathrel{\rlap{\lower4pt\hbox{\hskip1pt$\sim$}}
    \raise1pt\hbox{$>$}}}         
    
\documentclass[
    ,final            
  ]
  {aipproc}

\layoutstyle{6x9}

\begin{document}

\title{CIPANP 2009: Closing Talk}


\author{W. C. Haxton}{
  address={Institute for Nuclear Theory and Department of Physics, \\
  University of Washington, Seattle, WA 98195}
}

\begin{abstract}
 CIPANP 2009 is the tenth meeting of this series.  I look back at some of the key
 events of past meetings, comment on a few of the presentations of this meeting, and look
 foward to the next CIPANP gathering,  when first data from the LHC will be in hand.
 
\end{abstract}

\maketitle


\section{CIPANP Retrospective}
Alan Krisch, encouraged by Louis Rosen among others, began CIPANP, the Conference on the Intersections of Particle and Nuclear Physics, with a meeting held in Steamboat Springs,
Colorado, twenty-five years ago.  Every since, this meeting has been characterized 
by two constants,  great locations and interesting intersections --
the questions found at the boundaries of nuclear,
particle, and astrophysics, and the mix of theory, experiment, and instrumentation
needed to answer those questions.

As the San Diego meeting is a milestone, the tenth in the series and the
end of a quarter century of such efforts, I start
 this talk with a retrospective of past meetings.  I omit 
the ``travelogue" slides of the original, but the physics highlights remain:
\begin{itemize}
\item Steamboat Springs, 1984: CIPANP celebrated the discovery of the W 
and Z and the ground-breaking for LEP.
\item Lake Louise, 1986:  The MSW mechanism was
changing views of the solar $\nu$ problem, while claims for a 17 keV $\nu$
caused lively debates among experimentalists.
\item Rockport, Maine, 1988:   Theory papers on SN1987A greatly outnumbered the
$\nu$ burst events recorded by IMB and Kamiokande.
\item Tucson, 1991:  COBE data marked a transition into an era of precision
cosmology.
\item St. Petersburg, 1994:  We struggled
to reconcile the cancellation of the SSC with US aspirations to lead exploration of the
high-energy frontier.
\item Big Sky, Montana, 1997:  The top quark discovery gave us six quarks, and the newly inaugurated CEBAF was delivering first
beam.
\item Quebec City, 2000: Super-Kamiokande had announced the discovery
of $\nu$ mass, supernova data showed an expanding universe, and RHIC was
engaged in Run I.
\item New York, NY, 2003: WMAP year-one data, LIGO commissioning, and first results
from the Sudbury Neutrino Observatory were among the highlights.
\item Rio Grande, Puerto Rico, 2006:  RHIC finds a perfect fluid, and astronomers find
the first double pulsar.
\item San Diego, 2009:  Underground science facilities
(SNOLab opening, DUSEL site selection), ultra-high-energy cosmic rays,
and the start of Fermi's program to map the universe
in high-energy gamma rays are among the highlights.
\end{itemize}

As in baseball, I will try to be a good ``closer" by finishing quickly, while asking the question, 
what should we take away
from this meeting as ``homework" for 2012?\\
 
 \section{Inner Space, Outer Space}
 One of CIPANP's growing intersection areas is the inner space/outer space one,
 deep questions in particle physics that arise from cosmology and astrophysics:
 \begin{itemize}
 \item What is the dark matter and what role did it play in the evolution of large-scale structure?
 \item Why does the universe have a net baryon number, that is, an excess of baryons
 over antibaryons?  What determines the baryon-to-photon ratio?
 \item What are the mechanisms by which nature generates mass, and how can cosmology
 constrain unknowns such as the absolute scale of $\nu$ mass?
 \item What is dark energy?
 \end{itemize}
 Neutrino physics plays a role in the first three questions, while in the case of the fourth,
 it is curious that the dark energy density is $\sim m_\nu^4$.\\
 
 \noindent
 {\it Inner Space/Outer Space I: Neutrinos}~~
 The $\nu$ is unique among the standard-model (SM) fermions in lacking a charge or any other additively
 conserved quantum number.   The freedom from a conserved lepton number allows
 the $\nu$ both Majorana and Dirac mass terms, leading to a natural  explanation from the anomalous scale of $\nu$ masses, $m_{\nu_e}/m_e \lsim 10^{-6}$.
 On diagonalizing the seesaw mass matrix,
 one finds a light $\nu$ of mass $m_D (m_D/m_R)$.   For a heavy right-handed Majorana mass
 $m_R$, a natural suppression factor, $m_D/m_R $, emerges to explain
 why $\nu$s are so much lighter than other SM (Dirac) fermions.  
 If one associates the third-generation 
 $\nu$ mass with the atmospheric mass scale
 \[ m_\nu^{(3)} \sim \sqrt{m_{23}^\mathrm{atmos}} \sim 0.05 \mathrm{~eV} \sim m_D^{(3)} 
 \left( {m_D^{(3)} \over m_R} \right),\]
 and fixes $m_D^{(3)} \sim m_\mathrm{top~quark} \sim 180 \mathrm{~GeV}$, one finds $m_R \sim
 0.3 \times 10^{15} \mathrm{~GeV}$.  That is, the atmospheric 
 mass$^2$ splitting is consistent with an $m_R$
 very close to the unification scale of SUSY grand-unified theories, $\sim 10^{16}$ GeV.
 
 Several  key experimental challenges in $\nu$ physics are clear:
 \begin{itemize}
 \item  Demonstrate that there are no $\nu$ ``charges," that is, that lepton number (LN) is violated.
 The fortunate existence of
 even-even isotopes where the first-order weak decay $(A,Z) \rightarrow (A,Z+1) + e^- + \bar{\nu}_e$ is
 energetically forbidden allows sensitive searches for the second order,
 LN-violating decay $(A,Z) \rightarrow
 (A,Z+2) + 2e^-$.  Talks at this meeting
 described new experiments \cite{bb}, such as
 Majorana/GERDA, CUORE, EXO, and SNO+, that use large volumes
 and strive for excellent radiopurity and energy resolution 
 (important for distinguishing 0$\nu$
 $\beta \beta$ decay from the tail of the two-electron energy distribution of 
 the LN-conserving $2\nu$ process).
 \item  Determine the absolute scale of $\nu$ mass.   As oscillations
 test only mass differences $m_i^2-m_j^2$, this
 remains on open question. 
 Tritium $\beta$ decay provides the best current laboratory limit on the 
 $\nu_e$ mass, 2.2 eV, or equivalently
 $\sum_{i=1}^3 m_i \lsim 6.6$ eV.
 KATRIN
 could lower the $m_{\nu_e}$ mass limit to about 0.2 eV, and thus the bound on the sum
 of $\nu$ masses to $\sim 0.6$ eV.   But more stringent constraints may come from
 cosmological analyses \cite{cosmonu}: lighter $\nu$s remain relativistic longer, 
 travel further, and thus suppress the growth of structure for smaller wave
 numbers $k$ (larger distance scales),
 \[ k_\mathrm{free~streaming} \sim 0.004 \sqrt{m_\nu/0.05 \mathrm{eV}} ~\mathrm{Mpc}^{-1}. \]
 While cosmological analyses differ somewhat in their treatments of parameter
 correlations, typically limits are
  \[ \sum_{i=1}^3 m_i \lsim 0.7 \mathrm{~eV} \mathrm{~~or~equivalently~~}\rho_\nu \lsim 0.013 \rho_\mathrm{crit} \]
  where $\rho_\mathrm{crit}$ is the critical density that just closes the universe.  To ``measure"
  $\nu$ mass cosmologically at the lower bound determined by the atmospheric neutrino
  $m^2$ difference, 0.05 eV, ones needs a sensitivity to $\nu$ dark matter at $\sim$
  0.001 $\rho_\mathrm{crit}$.  This seems possible, given the advances anticipated in the
  next decade, including much more ambitious large-scale, high-Z structure surveys;
  additional constraints on large scales from improved CMB probes such as Planck;
  Lyman alpha forest focused on small scales and Z <  6; and weak lensing studies 
  probing medium to small scales -- provided, of course, that one knows how to
  combine data sets with somewhat different systematics.  
  \item Determine the mass hierarchy, and measure CP violation.   Two orderings of the
  mass eigenstates, normal and inverted, are consistent with oscillation 
  results.   This ambiguity could be resolved by a sufficiently precise cosmological measurement
  of the $\nu$ mass scale: if the cosmological mass is determined to be below
  0.10 eV, this would rule out the inverted hierarchy (as at least two eigenstates are
  then required to have masses $\gsim \sqrt{\delta m^\mathrm{atmos}} \sim 0.05$ eV).
         
  \hspace{0.5cm}A program of long baseline
  neutrino oscillation studies could  resolve questions about the $\nu$
  mass hierarchy as well as the size of $\nu$ CP violation.
  Matter effects -- the $\nu$ feels
  a flavor-dependent potential when it passes through the Earth -- depend on
  the hierarchy.   For long baselines, 1000-3000 km, the effects are large, and could be 
  explored with $\nu$ super beams couple to far detectors of mass $\gsim$ 100 ktons.
  In the case of a broad-band beam a distinctive oscillation
  pattern is imprinted on the spectrum, that allows one to separate the various effects of
  interest \cite{marciano}.
  
  \hspace{0.5cm} Long-baseline experiments are
  also sensitive to new sources of CP violation.  As other SM sources of CP violation
  appear too weak to account for the observed baryon number asymmetry, models in
  which the required CP violation resides among the leptons (leptogenesis) are in favor.
  A low-energy manifestation of this CP violation 
  would be phases in the $\nu$ mass matrix: three such phases exist in the
  three-generation case, one (Dirac) that
 could be measured in long-baseline  experiments and two (Majorana) that
 enter in processes like 0$\nu$ $\beta \beta$-decay.   The quantity
 that could be tested by, e.g.,  comparing
 $\nu_\mu \rightarrow \nu_\tau$ with $\bar{\nu}_\mu \rightarrow \bar{\nu}_\tau$,
 involves a product of $\nu$ mixing angles,
 all of which are large except for $\theta_{13}$.
 Unless $\theta_{13}$ is very small, there are
 long-baseline strategies (super beams for $\theta_{13} \gsim 10^{-2}$,  a $\nu$ factory for  $\gsim 10^{-4}$) for isolating
 the CP violation.  A demonstration that $\theta_{13}$ is nonzero in current reactor
 $\nu$ experiments, such as Double Chooz and Daya Bay, would spur efforts to mount
 long-baseline oscillation experiments.
 \end{itemize}
 
 This is an exciting science program:  $\beta \beta$ decay to determine whether the $\nu$ carries a
 LN, precision cosmology to determine the absolute scale of $\nu$ mass,
 and long-baseline experiments to measure CP violation and determine the mass
 hierarchy.  We need to get on with this program, now that we are eleven years into the massive $\nu$
 era.
 
 One concern I have about long-baseline $\nu$ physics concerns the nuclear
 physics.  Envisioned $\nu$ super beams 
 will peak at energies $\sim$ 2 GeV.  This is a
 complicated energy for the nuclear response, with both quasi-elastic and resonance
 contributions being important.  If we are constrained to do measurements at such energies,
 rather than in a deep inelastic regime of $\gsim$ 10 GeV,  adequate analysis
  tools must be developed.
  Even with calibration tests,
  it is difficult to envision an oscillation program achieving great precision in the absence
  of an adequate theoretical framework for parameterizing
  response functions.  Experimental analysis teams and nuclear theorists should be collaborating
  to develop the needed tools,
  including a program of validation against JLab electron scattering data.\\
  
  \noindent
  {\it Inner Space/Outer Space II: The High Energy Limits of the Universe}~~
  We heard wonderful talks on the importance of high energy/nuclear physics
  instrumentation to astrophysics efforts like Fermi/GLAST, IceCube, and
  Pierre Auger.   This is an exciting, emerging field that is testing the extremes
  of our universe.  Pierre Auger results suggest -- though the
  collaboration has not made a definite claim -- that the GZK cutoff
  (the threshold for protons to produce pions through interactions
  with cosmic microwave background (CMB) photons) is appearing at about $10^{19.5}$ eV.
  If the cosmos is opaque to ultra-high-energy (UHE) protons and nuclei, how will we 
  determine its high energy limits?
  
  Neutrinos propagate almost unaffected by matter or
  fields, and point back to their sources at cosmological distances.  
  The field has begun to develop very capable, large volume detectors for high energy
  $\nu$s.  IceCube, nearing completion, is focused on energies well below the GZK
  cutoff, where potential neutrino sources include AGNs and the explosive events in
  which gamma ray bursts are born.  IceCube complements cosmic ray observatories
  such as Pierre Auger: the km$^3$ volume was motivated by arguments that 
  connect the flux of $\nu$s to those of hadronic cosmic rays, assuming a
  source that is optically thin with respect to high energy proton-meson and 
  photo-meson interactions.  
  
  The existence of a GZK cutoff implies one source of UHE $\nu$s, the secondary
  produced in the decay of pions and neutrons that result from CMB photoproduction
  off cosmic ray (CR) protons and photo-dissociation of nuclei.  In addition to these GZK $\nu$s,
  there could be sources that, because of their extreme energies and radiation
  fields, are optically thin
  only to $\nu$s.  There may be "top-down" scenarios where super-energetic $\nu$s 
  are produced directly in the decays of exotic particles.  New methods under development
  to detect UHE $\nu$s include radio detection in ice and in the
  lunar limb, and fluorescence in the atmosphere.  One of the future challenges will
  be to use such methods to monitor very large detector volumes.
  
  The interactions of UHE CRs (and $\nu$s) with our atmosphere and with 
  ice and water targets involve center-of-mass energies significantly beyond the limits
  of terrestrial colliders like RHIC and the LHC.  The cascade codes developed to
  model such interactions -- including discriminating between UHE proton and
  nuclear collisions -- depend on extrapolations of laboratory data.  The uncertainties
  this induces in analyses is well appreciated \cite{engel}.  This is another example
  where close collaboration between the astrophysics and nuclear/particle physics
  communities may be important: the composition of UHE CRs is an important
  problem, affecting both our understanding of the sources and of CR 
  propagation through the CMB.\\
    
  \noindent
  {\it Inner Space/Outer Space III: Are We Done with Solar Neutrinos?}~~
  Here I mention a topic of personal interest, the prospect that
  future solar $\nu$ experiments might help us learn more about properties of the
  solar interior relevant to the Sun's very early history.  Recent improved analyses
  of photospheric absorption lines have led to a significant revision in abundances
  of volatile elements such as C, N, O, and Ne.  This has created a problem for the standard
  solar model (SSM), as sound speeds in the Sun's interior radiative zone are in good agreement
  with helioseismology only for the older abundances. Surface (photospheric lines)
  and interior (helioseismology) abundances are connected in the SSM through the
  assumption of a homogeneous zero-age-main-sequence (ZAMS) Sun: the
  protoSun is thought to have passed through a fully convective Hayashi phase.
  
  The discrepancy corresponds naively to a deficit in the 
  convective zone's total metal content of about 40 M$_\oplus$.  One can can speculate
  about mechanisms that might segregate metals at this level, subsequent to
  the Hayashi phase.
  Results from the Galileo and Cassini probes and
  planetary modeling show that significant metal differentiation occurred in the late-stage
  solar system disk.  Jupiter and Saturn are enriched
  ($\sim$ factor of four) in C, N, Ne, and similar elements, and the net giant-planet excess
  of metals is $\sim$ 40-90 M$_\oplus$.   This is thought to be a consequence of disk
  processes that concentrate larger grains and
  ice  in the disk's midplane ``dead zone" (where the rocky cores of the giant
  planets form) and metal-poor gas in the disk's outer layers.  Planetary formation occurs
  late in solar system evolution, after the protoSun is
  well formed, with  $\sim$ 5\% of the nebular gas remaining in the disk.  While midplane
  material is incorporated in planets, there are plausible mechanisms,
  including ionization of the surface by cosmic rays
  and x-rays, that could lead to deposition of the disk's surface gas onto the Sun.
  If the Sun has 
  developed a radiative core by this point, this could produce a two-zone sun with a
  convective zone relatively depleted in metals.
  
  Future solar neutrino experiments -- e.g., SNO+, a
  proposed larger, deeper version of Borexino -- may determine the metalicity of
  the core to an accuracy approaching $\sim$ 10\%.   This could be done by
  measuring the CN solar $\nu$s.
  The analysis \cite{serenelli} makes use
  of 1) the accurate measurements of the ${}^8$B neutrino flux by 
  Super-Kamiokande, which constrains the core temperature, and 2) recent
  progress in reducing uncertainties in the flavor
  physics and in the nuclear cross sections for the pp chain and CN cycle.
  
  \section{Low Energy/High Energy Intersections}
  Three subjects discussed frequently at this meeting -- CP violation, flavor physics,
  and dark matter --involve complementary efforts
  at the low-energy precision and high-energy intensity
  frontiers. \\
  
 \noindent
  {\it Low Energy/High Energy I: CP Violation}~~
  CP-violation was the theme of talks on
  low-energy searches for nonzero electric dipole moments, collider signals
  for supersymmetry, and the generation of a net baryon number
  through leptogenesis.
  
  Electric dipole moment (edm) experiments look for an interaction energy of the form
  \[ H_\mathrm{edm} = d~ \vec{E} \cdot \vec{s}, \]
  where $\vec{s}$ is a particle's spin. Because of the time reversal 
  properties $\vec{E}(t \rightarrow -t) \rightarrow \vec{E}$
  and $\vec{s}(t \rightarrow -t) \rightarrow -\vec{s}$, $H_\mathrm{edm}$ is 
  manifestly odd under $t \rightarrow -t$.
  One of the highlights of this meeting \cite{fortson} is shown in the last row of Table \ref{table:one},
  the recent factor-of-seven improvement in the edm of $^{199}$Hg, which previously
  competed with the neutron edm limit as the best constraint on a variety of sources
  of hadronic CP violation.  The Hg experiment was done in a vapor cell carefully
  prepared to minimize leakage currents (0.5-1.0 pA at 10 kV)
  and maximize the time for spin relaxation (100-200 s).  
  
  Over the next ten years significant progress is expected in this field.  The Hg experiment 
  might be improved by another factor $\sim$ 4 before leakage current limitations
  are reached.  New ultracold neutron experiments by groups at
  ILL, PSI, Munich and the SNS should improve neutron edm limits by about a factor 
  of 20, to $\sim 5 \times 10^{-28}$ e cm, by 2015.  The Princeton group is developing a new technique
  for measuring the edm of ${}^{129}$Xe in a high-density liquid state.  Techniques
  for measuring edms in traps are being developed for ${}^{213,225}$Ra and ${}^{223}$Rn
  at Argonne, KVI, and TRIUMF.  BNL is considering a
  proposal to measure the edms of deuterons circulating in a ring.
  
  Some of these methods will allow one to use systems where edms may be substantially
  enhanced, through level degeneracies or through nuclear collectivity.  The 5/2$^+$--5/2$^-$
  160 eV ground state parity doublet in ${}^{229}$Pa could produce an edm enhancement
  of $\sim$ 10,000.  A similar factor could arise in ${}^{225}$Ra, a nucleus where parity doublets
  arise from octupole deformation: in analogy with the more familiar quadrupole deformation
  in nuclei, the nucleus minimizes its energy in pear-shaped T-odd configurations, with the
  symmetry then restored by forming the even or odd combinations of these configurations.
  With new techniques like traps, one can use systems with
  nonzero atomic spins and higher nuclear spins: in a vapor cell, a nonzero atomic
  spin would lead to rapid loss of spin polarization, due to scattering off cell walls.  
  This opens up more opportunities to find enhancements and to
   probe higher-order T-odd nuclear moments such as the M2.  When one takes 
  into account the screening of a nuclear edm in a neutral atom, one finds that the 
  M2 response is enhanced by $R_A/R_N$, the ratio of atomic and nuclear sizes, relative
  to the C1 (edm) response.  (Part of this enhancement is lost because the M2 coupling is relativistic,
  but this suppression is not large in a heavy atom.)
  
\begin{table}[ht]
\begin{tabular}{| c | c | c | c |}
\hline 
Particle & edm limit (e cm) & system & SM prediction \\ \hline 
e & 1.9 $\times 10^{-27}$ & atomic $^{205}$Tl & $10^{-38}$ \\
p & 6.5 $\times 10^{23}$ & molecular TlF & $10^{-31}$ \\
n & 2.9 $\times 10^{-26}$ & ultracold n & $10^{-31}$ \\
${}^{199}$Hg & 2.1 $\times 10^{-28}$ $\rightarrow$ 3.1 $\times 10^{-29}$ & atom vapor cell & 10$^{-33}$ \\
\hline 
\end{tabular}
\caption{Electric dipole moment limits vs.  SM predictions for the
CKM phase.} 
\label{table:one}
\end{table}
 
Edm studies complement high energy efforts to find new sources of CP violation.
As identified SM sources of CP violation are not strong enough to 
account for the observed baryon number asymmetry, 
new sources of CP violation are expected.  We have already 
mentioned the phases in the $\nu$ mass matrix.  Extensions of the
SM, such as supersymmetry, are another very likely source
of new CP violation.  Making the connections between low-energy observables and fundamental
CP-violating phases requires significant theory.  In the case of edm studies of
diamagnetic atoms such as Hg or Ra, one is required to ``peel back" through layers
involving atomic screening and the CP-odd NN interaction, to get to quantities like the 
quark and squark edms and $\bar{\theta}$ that can be more readily related to the
fundamental CP-violating phases.  The prospect that the LHC may soon constrain
leading candidate theories that introduce new sources of CP violation, such as SUSY
theories, could greatly stimulate this field.\\

\noindent
{\it Low Energy/High Energy II: Flavor Physics}~~At this meeting we have heard a variety
of talks on flavor physics.  One of the fundamental questions in particle physics is
why we have three families.  At low energies new puzzles have emerged, such as the
origin of the large $\nu$ mixing angles (in contrast to the small
ones among quarks): to the extent we have been able to measure, $\theta_{23} \sim 45^\mathrm{o}$.

Table \ref{table:two} shows the current limits on a variety of lepton flavor violating (LFV) decays.
Facilities such as JPARC and FermiLab are considering high intensity, next-generation
experiments to significantly extend limits on $\mu \rightarrow e$ conversion.  The experiments
would make use of high-intensity pulsed proton beams to remove pion backgrounds by timing,
large acceptance capture solenoids to increase the useful muon flux, and bent solenoids to
transport muons while removing neutrals and separating charge.  The FermiLab experiment with
an 8 GeV proton beam could reach a branching ratio sensitivity of $\sim 4 \times 10^{-17}$, an
improvement of four orders of magnitude over current bounds.  This would,
for example, push sensitivities to tree-level LFV exchanges from the current mass of
$\sim$ 1 TeV to $\sim$ 10 TeV \cite{CERN} -- complementing the LHC's efforts to probe 
TeV-scale physics directly.  JPARC's 
experiment would use 40 GeV protons and might reach even further, to  $\sim 5 \times 10^{-19}$. \\

 \begin{table}[ht]
\begin{tabular}{| l | c | c | c |}
\hline 
Mode & Bound (90\% c.l.) & Year & Experiment/Lb  \\ \hline 
$\mu^+ \rightarrow e^- \gamma$ & $1.2 \times 10^{-11}$ & 2002 & MEGA/LAMPF  \\
$\mu^+ \rightarrow e^-e^+e^-$ & $1.0 \times 10^{-12}$ & 1988 & SINDRUM I/PSI  \\
$\mu^+e^- \leftrightarrow \mu^-e^+$ & $8.3 \times 10^{-11}$ & 1999 & PSI \\
$\mu^-\mathrm{Ti} \leftrightarrow e^-\mathrm{Ti}$ & $6.1 \times 10^{-13}$ & 1998 & SINDRUM II/PSI   \\
$\mu^-\mathrm{Ti} \leftrightarrow e^+\mathrm{Ca}^*$ & $3.6 \times 10^{-11}$ & 1998 & SINDRUM II/PSI \\
$\mu^-\mathrm{Pb} \leftrightarrow e^-\mathrm{Pb}$ & $4.6 \times 10^{-11}$ & 1996 & SINDRUM II/PSI \\
$\mu^-\mathrm{Au} \leftrightarrow e^-\mathrm{Au}$ & $7.0 \times 10^{-13}$ & 2006 & SINDRUM II/PSI \\
\hline 
\end{tabular}
\caption{Low-energy limits on branching ratios for LFV decays. References and
further details can be found in \cite{CERN}.} 
\label{table:two}
\end{table} 
 
 \noindent
 {\it Low Energy/High Energy III: Dark Matter}~~A third low energy/high energy interface is
 the nature of the dark matter.  This is perhaps the coming decade's greatest physics opportunity.
 It potentially unites some of our most exciting frontiers: finding SUSY at the LHC, 
 explaining the evolution of
 the large-scale structure of the universe, and developing ultra-low-background counting
 techniques for direct detection of dark matter at Gran Sasso, SNOLab, DUSEL, and other
 underground locations.  We know this problem is real and must involve beyond-the-SM
 new particles, perhaps the lightest SUSY particle.   Underground detection technologies
 now under development,
 such as
 cryogenic noble gas detectors, hopefully can be scaled to
 large volumes.  The problem being solved -- identifying the bulk of the matter in the cosmos --
 has the ``wow" factor.  
 
 \section{Theory, Modeling, and Computation}
 My last comments, on theory and computation, are inspired in part by the
 progress in lattice QCD and in cosmological/astrophysical modeling that we heard
 summarized at this meeting.  In lattice QCD, for example, NN phase shifts are now being
 calculated in fully dynamical QCD, and properties of the chiral phase transition are being
 determined in finite temperature calculations with pion masses near the physical value.
 The resources for computing are expanding, and new algorithms (including improved lattice
 actions) are making computation more efficient.
 
 Theoretical modeling is sometimes undervalued with respect to more traditional theory,
 which is focused on new concepts.  But in fact this difference is exaggerated, as
 numerical modeling is a testing ground for new theory and provides opportunities
 for exploring theory consequences.  The acceptance of and 
 critical role played by numerical modeling in cosmology and astrophysics in notable -- this young field
 grew up with high performance computing (HPC), and has easily integrated HPC
 into its core.  Similarly, particle and nuclear physics have many problems for which
 numerical simulation is the only way to quantitatively connect underlying theory to
 observation: examples include the properties of RHIC collisions, the nucleon form factors
 measured at JLab, and  the BaBaR and Belle searches for exotics.
 
 This ``editorial" is appropriate for CIPANP because computation is entering a new
 phase.  New machine technologies could increase the power
 of computation by a factor of 1000 over the next decade.  This would have
 extraordinary implications for the physical sciences, allowing
 rapid advancements in core nuclear/particle/astrophysics
 applications such as lattice QCD, core collapse supernove, and the formation of the 
 first stars.  Some quantities --one example is NN phase shifts -- will be calculable
 from fundamental theory to an accuracy that may surpass experiment.
  
 But these changes will require adjustments.  Machines will utilize  
 advanced architectures and be as costly as some of our flagship experimental facilities.
 A new style of collaborative computation will be needed, partnerships between
 physical scientists,
 applied mathematicians, and computational scientists
 to define the underlying mathematics of 
 physical processes, develop algorithms optimized to new architectures, and validate
 the results.  This is a new kind of intersection for our field -- one that if embraced,
 could make physics a leading discipline in computational research.
 
 \section{Conclusions}
 In three years, when we next meet at CIPANP, we will have entered the LHC era.  
 It should be an exciting time, with the high energy frontier taking center stage as 
 TeV-scale physics is revealed.
 
 Let me conclude by thanking our hosts, the CIPANP 2009 organizing committee, for 
 their efforts to bring us together in the attractive environment of San Diego. 
 Marvin Marshak, the organizing committee chair,  has invested a great deal of his
 time and energy to make this meeting a great success.  He was assisted by his 
 very able team, Dan Cronin-Hennessy, Priscilla Cushman, Joseph Kapusta, Peter
 Litchfield, Jeremy Mans, and Yong Qian.  We owe Marvin and his colleagues
 a hearty thanks for a job well done!
 


\begin{thebibliography}{9}

 \bibitem{bb}
 See, for example, the DNP/DPF/DAP/DPB Joint Study on the Future of NeutrinoPhysics,
 http://www.aps.org/policy/reports/multidivisional/neutrino/.
 
 \bibitem{cosmonu}
 S. Hannestad and Y.~Y.~Y. Wong, \emph{J. Cosmology and Astroparticle Phys.} \textbf{4}, 0707 (2007).
 
 \bibitem{marciano}
 M.~V. Diwan {\it et al.}, arXiv:hep-ex/0211001/ and \emph{Phys. Dev. D} \textbf{68}, 012002 (2003).
 
 \bibitem{engel}
 I.~C. Maris {\it et al.}, arXiv:0907:0409 (to be published in the proceedings of ISVHECRI 2008).
 
 \bibitem{serenelli} 
 W.~C. Haxton and A.~M. Serenelli, \emph{Ap. J.} \textbf{687}, 678 (2008).
 
 \bibitem{fortson}
 W.~C. Griffiths, M.~D. Swallows, T.~H. Loftus, M.~V. Romalis, B.~R. Heckel, and E.~N. Fortson,
 \emph{Phys. Rev. Lett.} \textbf{102}, 101601 (2009).
 
 
 \bibitem{CERN}
 M. Raidal {\it et al.}, \emph{Eur. Phys. J. C} \textbf{57}, 13 (2008).
 
 

\end{thebibliography}
\end{document}